\documentclass[prb,twocolumn,showpacs]{revtex4}

\usepackage{graphicx}
\usepackage{amsmath}
\usepackage{amsfonts}
\usepackage{bm}
\usepackage{bbm}

\begin{document}

\title{Influence of exciton-exciton correlations on the polarization characteristics of the polariton amplification in semiconductor microcavities}

\author{S. Schumacher}
\author{N. H. Kwong}
\author{R. Binder}

\affiliation{College of Optical Sciences,
             University of Arizona,
             Tucson, Arizona 85721, USA}

\date{\today}

\newcommand{\ME}{{\genfrac{}{}{0.0pt}{3}{\text{TM}}{\text{TE}}}}
\newcommand{\EM}{{\genfrac{}{}{0.0pt}{3}{\text{TE}}{\text{TM}}}}
\newcommand{\EE}{{\genfrac{}{}{0.0pt}{3}{\text{TE}}{\text{TE}}}}
\newcommand{\MM}{{\genfrac{}{}{0.0pt}{3}{\text{TM}}{\text{TM}}}}
\newcommand{\TM}{\text{TM}}
\newcommand{\TE}{\text{TE}}
\newcommand{\probe}{{\,\sigma_s}}
\newcommand{\pump}{{\,\sigma_p}}
\newcommand{\probepump}{{\,{{\sigma_s}},{{\sigma_p}}}}

\newcommand{\XY}{{\genfrac{}{}{0.0pt}{3}{\text{X}}{\text{Y}}}}
\newcommand{\YX}{{\genfrac{}{}{0.0pt}{3}{\text{Y}}{\text{X}}}}
\newcommand{\X}{\text{X}}
\newcommand{\Y}{\text{Y}}


\begin{abstract}
Based on a microscopic many-particle theory we investigate the
influence of excitonic correlations on the vectorial polarization
state characteristics of the parametric amplification of polaritons
in semiconductor microcavities. We study a microcavity with perfect
in-plane isotropy. A linear stability analysis of the cavity
polariton dynamics shows that in the co-linear (TE-TE or TM-TM)
pump-probe polarization state configuration, excitonic correlations
diminish the parametric scattering process whereas it is enhanced by
excitonic correlations in the cross-linear (TE-TM or TM-TE)
configuration. Without any free parameters, our microscopic theory
gives a quantitative understanding how many-particle effects can
lead to a rotation or change of the outgoing (amplified) probe
signal's vectorial polarization state relative to the incoming
one's.
\end{abstract}

\maketitle

\section{Introduction}

In the past decade the parametric amplification of polaritons in
planar semiconductor microcavities has been the subject of intense
experimental and theoretical research, see, e.g.,
Refs.~\onlinecite{Savvidis2000,Huang2000,Ciuti2000,Stevenson2000,Saba2001,Whittaker2001}
or the reviews given in
Refs.~\onlinecite{Ciuti2003,Baumberg2005,Keeling2007}. In a typical
pump-probe setup in a co-circular polarization configuration the
amplification of a weak probe pulse has mainly been attributed to
four-wave mixing (FWM) processes mediated by the repulsive Coulomb
interaction of the exciton constituent of the polaritons excited on
the lower polariton branch
(LPB).\cite{Ciuti2000,Stevenson2000,Whittaker2001,Savasta2003} For a
specific pump in-plane momentum (defining the so-called ``magic
angle''), energy and momentum conservation is best fulfilled for the
FWM processes and thus a pronounced angular dependence of this
amplification is observed.\cite{Savvidis2000,Ciuti2000} Since in the
strong coupling regime the LPB is spectrally well below the
two-exciton scattering continuum, the influence of excitonic
correlations in the scattering processes of polaritons on the LPB is
strongly suppressed [compared to the situation in a single quantum
well (QW) without the strong coupling to a confined photon cavity
mode\cite{Schumacher2006}]. However, even for co-circular pump-probe
excitation these correlations must be considered for a complete
understanding of the experimental
results.\cite{Saba2001,Savasta2003,Savasta2003b}

Whereas for co-circular pump-probe excitation only exciton-exciton
scattering in the electron-spin triplet channel plays a role, for
excitations in other vectorial polarization state configurations,
excitonic scattering in the electron-spin singlet channel is also
expected to contribute to the amplification mechanism. In the latter
case, a change (in the following loosely referred to as `rotation')
of the vectorial polarization state of the amplified probe signal
compared to the incoming one's can be attributed to this coupling of
the two spin subsystems excited with right ($+$) and left ($-$)
circularly polarized light, respectively.
\cite{Bolger1996,Ramkumar2000,Lagoudakis2002,Kavokin2003,Eastham2003,Renucci2005,Kavokin2005,Dasbach2005,Krizhanovskii2006}
However, different effects can overshadow rotations in the vectorial
polarization state that are caused by the spin-dependent
many-particle interactions that mediate the amplification process,
e.g., a splitting of the TE and TM cavity
modes\cite{Panzarini1999,Kavokin2004} (longitudinal-transverse
splitting), or an in-plane anisotropy of the embedded QW or the
cavity\cite{Krizhanovskii2006,Romanelli2007}. Furthermore, for not
linearly polarized pump excitation an imbalance in the polariton
densities in the two spin subsystems ($+$ and $-$) will also lead to
a rotation in the vectorial polarization state of the amplified
signal.\cite{Eastham2003,Krizhanovskii2006} These different
mechanisms have previously been
investigated\cite{Eastham2003,Kavokin2003,Kavokin2005,Renucci2005,Krizhanovskii2006}
based on models describing the effective polariton dynamics in the
cavity. In these models that describe the system dynamics at the
polariton quasi-particle level, the spin-dependent
polariton-polariton scattering matrix elements are included as input
parameters for the theory. With a reasonable choice of the parameter
set, good agreement with experimental results showing rotations in
the probe's vectorial polarization state has been
obtained.\cite{Kavokin2003,Kavokin2005,Renucci2005,Krizhanovskii2006}

In contrast to these previous
studies,\cite{Eastham2003,Kavokin2003,Kavokin2005,Renucci2005,Krizhanovskii2006}
we employ a microscopic theory that calculates, from a few material
parameters, the scattering matrices driving the polariton
amplification in the different vectorial polarization state
channels\cite{Takayama2002}. No additional assumptions for the
effective polariton-polariton interaction are needed, which is
directly included in our theory via the frequency dependent and
complex exciton-exciton scattering matrices.\cite{Savasta2003} Our
theoretical analysis incorporates: (i) the well-established
microscopic many-particle theory for the optically-induced QW
polarization dynamics based on the dynamics-controlled truncation
(DCT) formalism\cite{Axt1994a,Oestreich1995}, and (ii) the
self-consistent coupling of this dynamics to the dynamics of the
optical fields in the cavity
modes\cite{Kwong2001a,Kwong2001b,Savasta2003} including all
vectorial polarization state channels. The theory consistently
includes all coherent third order ($\chi^{(3)}$) nonlinearities and
the resulting equations of motion are solved in a self-consistent
fashion in the optical fields which includes a certain class of
higher-order
nonlinearities.\cite{Oestreich1999,Buck2004,Schumacher2005a}
Correlations involving more than two excitons and those involving
incoherent excitons are neglected. These effects are not expected to
qualitatively alter the presented results for the considered
coherent exciton densities of $\sim10^{10}\,\mathrm{cm^{-2}}$,
especially for excitation well below the exciton resonance.

Based on this theory we introduce a linear stability analysis (LSA)
of the cavity polariton dynamics as a general and powerful tool to
study the role of spin-dependent polariton-polariton scattering
(including time-retarded quantum correlations). For steady-state
pump excitation and as long as depletion of the pump from scattering
into probe and FWM signals can be neglected, the LSA gives
comprehensive information about growth and/or decay (in the
following only referred to as `growth') rates for probe and FWM
intensities in all vectorial polarization states. The growth rates
determine the exponential growth of components in different
vectorial polarization states over time, and determine together with
the initial conditions the ratio of these components after a given
growth duration, uniquely determining the final vectorial
polarization state. Although the results are obtained for strict
steady-state pump excitation, as discussed below they can to a large
extent be carried over to the analysis of pump-probe experiments
with finite pulse lengths.\cite{Schumacher2006}

We use this theory to investigate a microcavity system with perfect
in-plane isotropy. As an intrinsic effect that is not caused by
structural imperfection, we include a splitting of the TE and TM
cavity modes as shown in Fig.~\ref{FigTmatrices}(a). This way we
study a system where all vectorial polarization state rotations of
the amplified probe signal can unambiguously be traced back to
intrinsic phenomena always present in planar semiconductor
microcavities, the TE-TM cavity-mode splitting and the
spin-dependent polariton-polariton scattering mediating the
amplification.

For this system we analyze results that show how for a linearly
polarized pump many-particle correlations and the TE-TM cavity-mode
splitting lead to different growth rates of the linearly polarized
components (TE,TM) in the probe pulse. This difference can lead to
rotation in the vectorial polarization state of the amplified probe
compared to the incoming
one's.\cite{Eastham2003,Kavokin2003,Kavokin2005,Renucci2005,Krizhanovskii2006}
Starting from the equations governing the cavity-polariton dynamics,
we take advantage of our theoretical approach to isolate and discuss
the frequency-dependent scattering matrices that give rise to this
difference in the vectorial polarization state channels. We show
that in the studied regime, close to the amplification threshold,
even where the correlation contribution in the spin-singlet channel
is weak\cite{Lagoudakis2002,Dasbach2005}, these correlations can
give rise to an almost complete vectorial polarization state
rotation into the ``preferred'' cross-linear (TE-TM or TM-TE)
pump-probe configuration for the pump and outgoing probe pulses. For
sufficiently long amplification duration, this result can become
virtually independent of the input probe's vectorial polarization
state as long as it contains a small component polarized
perpendicular to the linearly polarized pump.

\section{The theoretical model}\label{sectheory}

\begin{figure}[t]
\includegraphics[scale=0.85]{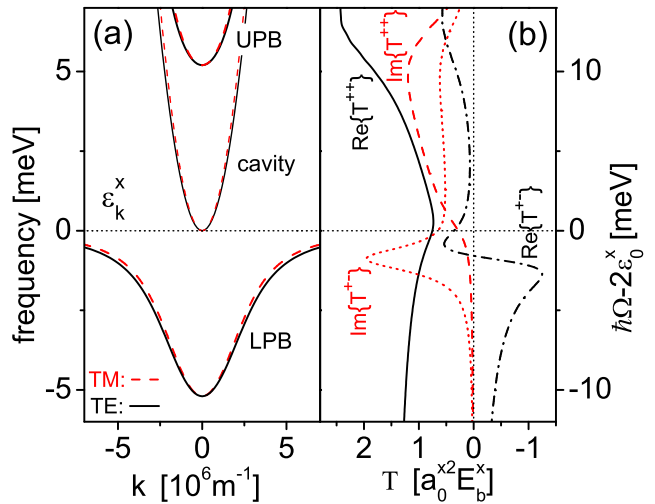}
\caption{\label{FigTmatrices} (color online) (a) Shown is the
dependence of the cavity polariton modes on the magnitude $k$ of the
in-plane momentum. Depicted are the lower (LPB) and upper (UPB)
polariton branches for TE (solid) and TM (dashed) cavity modes, and
the bare cavity and exciton (dotted) dispersions. For details on the
modeling of the cavity modes see Sec.~\ref{sectheory}. (b) Real and
imaginary parts of the two-exciton scattering matrices
$\tilde{\mathcal{T}}(\Omega)$ in the co-circular ($++$) and
counter-circular ($+-$) polarization state channels. (a,b) Results
are shown for typical GaAs parameters\cite{CavityFootnote1} as used
throughout this work.}
\end{figure}

We use a microscopic many-particle theory to describe the coherent
QW response to the light field confined in the cavity. Based on the
dynamics-controlled truncation (DCT) approach
\cite{Axt1994a,Oestreich1995} all coherent optically-induced third
order nonlinearities, i.e., phase-space-filling (PSF), excitonic
mean-field (Hartree-Fock) Coulomb interaction and two-exciton
correlations are included on a microscopic level. We use a two-band
model (including spin-degenerate conduction and heavy-hole valence
band) to describe the optically induced polarization in the GaAs
QW.\cite{CavityFootnote1} Since we are mainly interested in pump
excitation in the LPB, i.e., energetically below the bare exciton
resonance (cf. Fig.~\ref{FigTmatrices}), we account for the dominant
contributions to the QW response by evaluating the optically induced
QW polarization in the 1s heavy-hole exciton
basis.\cite{Donovan2001,Kwong2001b,Buck2004,Schumacher2005a}

\begin{figure}[t]
\includegraphics[scale=0.9]{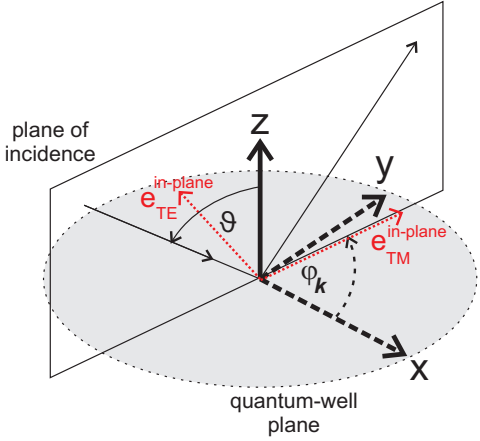}
\caption{\label{schematic} (color online) Schematic of the
excitation geometry. The plane of incidence is spanned by the wave
vector of the incoming light field and the $z$ axis. All lines in
this plane are solid. The quantum-well plane is the x-y plane. All
lines in this plane are broken. The figure shows the basis vectors
$\mathbf{e}_{\TE}^{\text{in-plane}}$ and
$\mathbf{e}_{\TM}^{\text{in-plane}}$ that span the projection of the
TE-TM basis on the x-y plane.  The polar angle $\vartheta$ and
azimuthal angle $\varphi_{\mathbf{k}}$ are also shown. For more
explanation see text in Sec.~\ref{sectheory}.}
\end{figure}

We start from the coupled equations of motion for the field
$\mathbf{E}_\mathbf{k}$  in the cavity modes with in-plane momentum
$\mathbf{k}$ (treated in quasi-mode approximation\cite{Savasta1996})
and the optically induced interband polarization amplitude
$p_\mathbf{k}$ in the embedded QW. We formulate our theory in the
TE-TM basis for the optical fields in the cavity,
$\mathbf{E}_\mathbf{k}={E}_\mathbf{k}^{\TE}\mathbf{e}_{\TE}+{E}_\mathbf{k}^{\TM}\mathbf{e}_{\TM}$
(see Fig.~\ref{schematic} for the excitation geometry), where the
field components in the TE mode
${E}_\mathbf{k}^{\TE}\mathbf{e}_{\TE}$ (also called {s-polarized})
are characterized by an electric field vector with in-plane (in the
plane of the QW) component perpendicular to the in-plane momentum
$\mathbf{k}$, and field components in the TM mode
${E}_\mathbf{k}^{\TM}\mathbf{e}_{\TM}$ (also called {p-polarized})
by an electric field vector with in-plane component parallel
$\mathbf{k}$. In this basis it is most intuitive to include
different cavity-mode exciton couplings for the TE and TM modes in
the theory: the in-plane component of the fields in the TE mode does
not depend on the polar angle of incidence $\vartheta$ (cf.
Fig.~\ref{schematic}) and thus the coupling strength to the
excitonic dipole in the quantum-well plane does not depend on
$\vartheta$. A different result is found for fields in the TM mode
where the magnitude of the in-plane component depends on the polar
angle of incidence ${\vartheta}$. Since the $z$ component of fields
in the TM mode does not couple to the excitonic dipole for
excitation of heavy-hole excitons in the QW\cite{Savona1996}, the
effective coupling constant of excitons and fields in the TM mode
decreases with the polar angle like $\sim\cos{\vartheta}$.
Additionally, we include a slightly different polar angular
dependence of the bare TE and TM cavity dispersions $\omega_k^{\ME}$
that in general depend on the specific materials and design of the
cavity.\cite{Panzarini1999} The resulting cavity dispersions are
shown in Fig.~\ref{FigTmatrices}(a) and the parameters will be given
later in this section. In order not to complicate the structure of
the nonlinear terms in the equations of motion for the polarization
amplitudes, we use the usual Cartesian basis (X-Y basis) in the QW
plane ($x-y$ plane) to decompose the polarization into its
components as
$\mathbf{p}_{\mathbf{k}}={p}_{\mathbf{k}}^\X\mathbf{e}_x+{p}_{\mathbf{k}}^\Y\mathbf{e}_y$.
In the X-Y basis, the projection of the TE-TM basis vectors on the
$x-y$ plane, $\mathbf{e}_\TE^{\text{in-plane}}$ and
$\mathbf{e}_\TM^{\text{in-plane}}$, rotates with the in-plane
component $\mathbf{k}$ of the momentum of the incident wave. The
azimuthal angle between $\mathbf{k}$ and the $x$ axis is denoted by
$\varphi_{\mathbf{k}}$ (cf. Fig.~\ref{schematic}). Simple geometric
considerations lead to the azimuthal angular dependencies that
appear in the terms coupling the equations of motion for field and
polarization amplitude components in the different bases:
\begin{multline}
\label{fieldmotion}
i\hbar\dot{E}^{\ME}_\mathbf{k}=\Big(\hbar\omega_k^{\ME}-i\gamma_c\Big)E_\mathbf{k}^{\ME}
\\
-V_{{k}}^{\ME}\Big[p_\mathbf{k}^\XY\cos\varphi_{\mathbf{k}}\pm
p_\mathbf{k}^{\YX}\sin{\varphi_{\mathbf{k}}}\Big]+i\hbar
t_cE_{\mathbf{k},\text{inc}}^{\ME}\,,
\end{multline}
\begin{widetext}
\begin{align}\label{pumpmotion}
i\hbar&\dot{p}_\mathbf{k}^\XY=\Big(\hbar\omega^x_{k}-i\gamma_x\Big)p^\XY_\mathbf{k}-\Big(V_{{k}}^\ME E_\mathbf{k}^\ME\cos{\varphi_{\mathbf{k}}} \mp V_{{k}}^\EM E_\mathbf{k}^\EM\sin{\varphi_{\mathbf{k}}}\Big) \nonumber \\
+&\tilde{A}\sum_{\mathbf{k}'\mathbf{k}''}\Big(p^{\XY\ast}_{\mathbf{k}'+\mathbf{k}''-\mathbf{k}}p^{\XY}_{\mathbf{k}'}+p^{\YX\ast}_{\mathbf{k}'+\mathbf{k}''-\mathbf{k}}p^{\YX}_{\mathbf{k}'}\Big)\Big(V_{{k}''}^\ME
E_{\mathbf{k}''}^\ME\cos{\varphi_{\mathbf{k}''}} \mp V_{{k}''}^\EM
E_{\mathbf{k}''}^\EM\sin{\varphi_{\mathbf{k}''}}\Big) \nonumber \\
+&\tilde{A}\sum_{\mathbf{k}'\mathbf{k}''}\Big(p^{\YX\ast}_{\mathbf{k}'+\mathbf{k}''-\mathbf{k}}p^{\XY}_{\mathbf{k}'}-p^{\XY\ast}_{\mathbf{k}'+\mathbf{k}''-\mathbf{k}}p^{\YX}_{\mathbf{k}'}\Big)\Big(V_{{k}''}^\EM
E_{\mathbf{k}''}^\EM\cos{\varphi_{\mathbf{k}''}} \pm V_{{k}''}^\ME
E_{\mathbf{k}''}^\ME\sin{\varphi_{\mathbf{k}''}}\Big) \nonumber \\
+&\frac{1}{2}\sum_{\mathbf{k}'\mathbf{k}''}
p_{\mathbf{k}'+\mathbf{k}''-\mathbf{k}}^{\XY\ast}\int_{-\infty}^{\infty}{\mathrm{d}\,t'\Big[\big(\mathcal{T}^{++}(t-t')+\mathcal{T}^{+-}(t-t')\big)p_{\mathbf{k}'}^\XY(t')
p_{\mathbf{k}''}^\XY(t')}-\big(\mathcal{T}^{++}(t-t')-\mathcal{T}^{+-}(t-t')\big)p_{\mathbf{k}'}^\YX(t')
p_{\mathbf{k}''}^\YX(t')\Big] \nonumber \\
+&\sum_{\mathbf{k}'\mathbf{k}''}
p_{\mathbf{k}'+\mathbf{k}''-\mathbf{k}}^{\YX\ast}\int_{-\infty}^{\infty}{\mathrm{d}\,t'\,\Big[\mathcal{T}^{++}(t-t')p_{\mathbf{k}''}^\XY(t')p_{\mathbf{k}'}^\YX(t')
}\Big] \,.
\end{align}
\end{widetext}
These equations constitute the generalization of the equations given
in Ref.~\onlinecite{Savasta2003}, now including all vectorial
polarization states. As discussed above, the polarization amplitudes
are given in the X-Y basis, while the fields are given in the TE-TM
basis.\cite{CavityFootnote2} The meaning of the symbols in
Eq.~(\ref{pumpmotion}) are to be discussed in the remainder of this
paragraph, along with the used parameters and approximations.
Unless otherwise noted, the time argument in
Eqs.~(\ref{fieldmotion}) and (\ref{pumpmotion}) is $t$. $t_c$ is the
coupling constant of the cavity mode to the external light fields
$E_{\mathbf{k},\text{inc}}^{\ME}$, and the dephasing constant
$\gamma_c$ describes optical losses from the cavity to the outside
world.\cite{Savasta2003} The dependence of the bare cavity modes
$\omega^{\ME}_k$ and the dependence of the exciton-cavity mode
coupling $V_k^\ME$ on the polar angle $\vartheta$ is modeled on a
phenomenological level along the guidelines given in
Ref.~\onlinecite{Panzarini1999}. We approximate the bare cavity
dispersions with
$\omega^{\text{TM}}_k=\frac{\omega_0}{\cos{\vartheta}}+100\,\mathrm{meV}\cdot\sin^2\theta$
and $\omega^{\text{TE}}_k=\frac{\omega_0}{\cos{\vartheta}}$ with
$\sin\vartheta=\frac{|\mathbf{k}|c_0}{\omega{n_{\text{bg}}}}$. This
way a TE-TM cavity-mode splitting from the mismatch of the center of
the stopband of the cavity mirrors and the Fabry-Perot frequency of
the cavity is phenomenologically included.\cite{Panzarini1999} The
cavity-mode exciton couplings are
$V_{k}^{\text{TM}}=V_{0}^{\text{TM}}\cos\vartheta$ and
$V_{k}^{\text{TE}}=V_{0}^{\text{TE}}$, respectively, with
$V_{0}^{\text{TM}}=V_{0}^{\text{TE}}=5.2\,\mathrm{meV}$. With
$\hbar\omega_0^\ME=\varepsilon^x_0$ we assume zero cavity-mode
exciton splitting for $k=0$. The chosen
parameters\cite{CavityFootnote1} give a reasonable magnitude and
polar angular dependence of the TE-TM mode splitting in the LPB,
comparable to the results in, e.g., Ref.~\onlinecite{Kavokin2004}.
Since the presented results do not crucially depend on the details
of the cavity-mode splitting, no further insight is expected from a
more elaborate treatment. In Eq.~(\ref{pumpmotion}),
$\hbar\omega^x_k$ is the 1s heavy-hole exciton in-plane dispersion,
$\gamma_x$ a phenomenological dephasing constant of the excitonic
polarization amplitude, and $\tilde{A}$ is related to the excitonic
PSF constant $A^{\text{PSF}}$ by
$\tilde{A}=\frac{A^{\text{PSF}}}{\phi_{1s}^{\ast}(\mathbf{0})}$,
with $\phi_{1s}(\mathbf{r})$ being the two-dimensional QW exciton
wavefunction. Without loss of generality, in the following the
quantities $\phi_{1s}$, $V_0^\ME$, and $t_c$ are chosen to be
real-valued. The parameter values are listed in
Ref.~\onlinecite{CavityFootnote1}. Although in this paper we
investigate a spatially isotropic microcavity-system, spatial
anisotropy can easily be included in the theory via
$\omega_k^x\rightarrow\omega_{\mathbf{k}}^x$ to model an anisotropic
dispersion of the QW excitons and via
$\omega_k^\ME\rightarrow\omega_{\mathbf{k}}^\ME$ to model anisotropy
of the cavity modes. The two-exciton scattering matrices
(T-matrices) $\mathcal{T}$ in the co-circular ($++$) and
counter-circular ($+-$) polarization state channels include a
two-exciton dephasing rate $2\gamma$ and are given by
$\mathcal{T}^{++}=\mathcal{T}^+$ and
$\mathcal{T}^{+-}=(\mathcal{T}^++\mathcal{T}^-)/2$, with the
T-matrices $\mathcal{T}^+$ and $\mathcal{T}^-$ in the electron-spin
triplet and singlet channel, respectively, as defined in Eq.~(32) of
Ref.~\onlinecite{Takayama2002}. The frequency dependence of real and
imaginary parts of $\mathcal{T}^{++}$ and $\mathcal{T}^{+-}$ is
shown in Fig.~\ref{FigTmatrices}(b). We neglect the momentum
dependence of the T-matrices for scattering processes involving two
excitons with different in-plane momenta. Calculations in a
different context have shown that this is justified in a good
approximation for the small optical momenta contributing
here.\cite{Kwong2007} We also neglect possible corrections to the
excitonic T-matrices from the coupling to the photons in the cavity
modes. This is supported by experimental observations that indicate
that even in the strong coupling regime the biexciton binding energy
is not significantly affected by the coupling to the cavity
modes.\cite{Borri2000} Also, good theory-experiment agreement has
been achieved in
Refs.~\onlinecite{Kuwatagonokami1997,Kwong2001a,Kwong2001b} using a
theory based on pure exciton-exciton scattering matrices.

To simulate a typical pump-probe setup, we start from
Eqs.~(\ref{fieldmotion}) and (\ref{pumpmotion}) and chose a finite
in-plane momentum $\mathbf{k}_p$ for the pump propagating along a
given axis, here, without loss of generality the $x$ axis, i.e.,
$\varphi_{\mathbf{k}_p}=0$. We ``detect'' the probe in normal
incidence with in-plane momentum $\mathbf{k}=0$ where in the past
the strongest amplification has been
observed.\cite{Kavokin2003,Savasta2003,Savasta2003b,Gippius2004}
This fixes the in-plane momentum of the background-free FWM signal
to $2\mathbf{k}_p$. We go beyond an evaluation of the theory on a
strict $\chi^{(3)}$ level by self-consistently calculating the
resulting excitonic and biexcitonic polarization amplitude dynamics
up to arbitrary order in the pump field\cite{Oestreich1999,Buck2004}
and we linearize the equations of motion in the weak probe field.
Only via this self-consistent solution the coupling of the probe
signal to the background-free FWM signal is included in the theory
which provides the basic feedback mechanism that leads to the
unstable behavior in, e.g.,
Refs.~\onlinecite{Savvidis2000,Huang2000,Ciuti2000,Stevenson2000,Saba2001,Ciuti2003,Baumberg2005}.
We limit our analysis to coherent exciton densities that are low
enough [$\lesssim2\times10^{10}\,\mathrm{cm^{-2}}$, cf.
Figs.~\ref{Fig2} and \ref{Fig3}] so that neglect of higher than
two-exciton Coulomb correlations can be justified.

\section{Linear stability analysis}\label{Secstabana}

\begin{figure}[t]
\begin{center}
\includegraphics[scale=0.98]{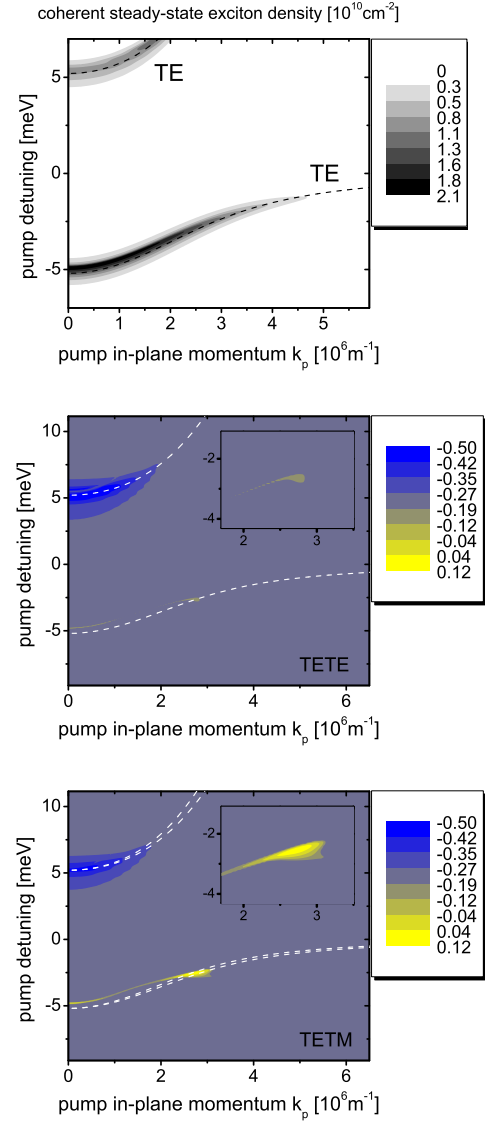}
\end{center}
\caption{\label{Fig2}(color online) Top: Coherent steady-state
exciton density $|p_{k_p}^\Y|^2$ for a fixed pump intensity of a
linearly TE polarized pump as a function of the magnitude $k_p$ of
the pump in-plane momentum and the pump detuning
$\Delta\varepsilon=\hbar\omega_p-\varepsilon^x_0$ from the bare
exciton resonance $\varepsilon^x_0$. The linear polariton
dispersions are included as the dashed lines. Middle and lower
figures show the real part of the eigenvalue of
$M^{\TE,\TE}_\parallel$ and $M^{\TE,\TM}_\bot$ with the largest real
part in $\mathrm{meV}$ (maximum growth rate if larger than zero).
The linear polariton dispersions are included as the dashed lines
and the insets show the same data around the ``magic angle'' for
pump excitation on the LPB.}
\end{figure}

\begin{figure}[t]
\begin{center}
\includegraphics[scale=0.98]{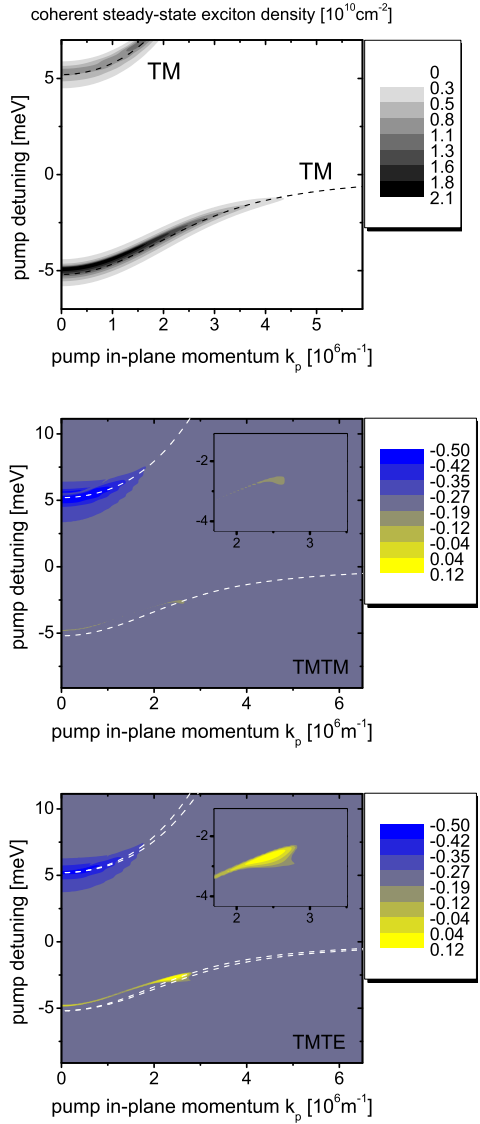}
\end{center}
\caption{\label{Fig3} (color online) Same as Fig.~\ref{Fig2} but for
linearly TM polarized pump.}
\end{figure}

In this section we introduce the linear stability analysis (LSA)
used in the remainder of this paper. To analyze the stability of the
pump-probe dynamics the LSA is done without an incoming probe field
and for a linearly polarized monochromatic continuous wave (cw) pump
field
$E^\ME_{\mathbf{k}_p}(t)=\tilde{E}^\ME_{\mathbf{k}_p}(\omega_p)\mathrm{e}^{-i\omega_pt}$
inducing the pump polarization amplitude
$p^\XY_{\mathbf{k}_p}(t)=\tilde{p}^\XY_{\mathbf{k}_p}(\omega_p)\mathrm{e}^{-i\omega_pt}$,
with $\dot{\tilde{p}}^\XY_{\mathbf{k}_p}=0$ and
$\dot{\tilde{E}}^\ME_{\mathbf{k}_p}=0$ ($\omega_p$ is the pump
frequency). The pump polarization amplitude is a solution of the
cubic nonlinear pump equation following from
Eqs.~(\ref{fieldmotion}) and (\ref{pumpmotion}) for unidirectional
light propagation and is determined as outlined in
Appendix~\ref{AppA}. The resulting coherent exciton density
$|p^\XY_{\mathbf{k}_p}|^2$ for excitation with a linearly polarized
pump of fixed intensity is shown in Figs.~\ref{Fig2} and \ref{Fig3}
as a function of the magnitude of the pump in-plane momentum
$|\mathbf{k}_p|$ and pump detuning $\Delta\varepsilon$ from the bare
exciton resonance for excitation in the TE or TM mode, respectively.
No bistable behavior of the pump-induced exciton density in
Figs.~\ref{Fig2} and \ref{Fig3} is found (this follows from the
solution of the nonlinear pump equation as outlined in
Appendix~\ref{AppA}). However, for other values of cavity or QW
parameters and a different pump intensity, bistability may occur,
which would complicate our discussion\cite{Gippius2004,Wouters2007}.
The pump densities shown have their maxima close to the linear
polariton dispersions (included as the dashed lines) and decrease
along the LPB with increasing in-plane momentum because of the
decreased coupling of the (mostly exciton-like) large-momentum
polariton states to the incoming field. Furthermore, a reduced
exciton density is found for excitation of the UPB caused by strong
excitation-induced dephasing (EID) for pump excitation in the
two-exciton scattering continuum (the spectral region where the
two-exciton scattering matrices shown in Fig.~\ref{FigTmatrices}
exhibit a large imaginary part). For the stability analysis we
evaluate the memory integrals in Eq.~(\ref{pumpmotion}) contributing
to the probe and FWM directions in a Markov approximation for the
two-exciton scattering continua in the T-matrices $\mathcal{T}^{++}$
and $\mathcal{T}^{+-}$. Our Markov approximation is effected by
taking $p_{0}(t')\approx p_{0}(t)\mathrm{e}^{i\omega_p(t-t')}$ and
$p_{2k_p}(t')\approx p_{2k_p}(t)\mathrm{e}^{i\omega_p(t-t')}$ where
the probe and FWM polarization amplitudes $p_{0}$ and $p_{2k_p}$
appear under the time-retarded integrals in Eq.~(\ref{pumpmotion})
together with the continuum contributions in $\mathcal{T}^{++}$ and
$\mathcal{T}^{+-}$. In contrast, we include the bound biexciton
state exactly via the time-dependent amplitudes
$b_{\mathbf{k}_p}(t)$ and $b_{3\mathbf{k}_p}(t)$. For this we
separate the bound biexciton contributions $\mathcal{T}^{+-}_{xx}$
from the correlation kernels $\mathcal{T}^{+-}$ in
Eq.~(\ref{pumpmotion}) as
$\mathcal{T}^{+-}=\mathcal{T}^{+-}_{\text{cont}}+\mathcal{T}^{+-}_{xx}$.
The bound biexciton contributions to Eq.~(\ref{pumpmotion}) can be
exactly included via the equations of motion [cf. Eqs.~(11) and (12)
of Ref.~\onlinecite{Takayama2002}] for the biexciton amplitudes
$b_{\mathbf{k}_p}(t)$ and $b_{3\mathbf{k}_p}(t)$ which are labeled
according to the total in-plane momentum of their source terms:
$\sim p_0p_{\mathbf{k}_p}$ and $\sim
p_{2\mathbf{k}_p}p_{\mathbf{k}_p}$, respectively. This way we
include quantum memory effects related to the excitation of bound
biexcitons, which were previously shown to play an important role in
the study of FWM instabilities in single semiconductor QWs
\cite{Schumacher2006}. Since, for the chosen excitation geometry
with finite pump in-plane momentum $\mathbf{k}_p$, the probe and FWM
signals do not oscillate at the pump frequency $\omega_p$, the
Markov approximation for the two-exciton scattering continuum may
not be as justified as it is for the single QW system investigated
in Ref.~\onlinecite{Schumacher2006}. However, close to or in the
unstable regime, those wave mixing processes that describe the
pairwise scattering of pump polaritons into the probe and FWM
directions play the dominant role in the probe and FWM dynamics. For
monochromatic cw pump excitation these terms are of purely Markovian
nature and hence the T-matrices in these terms contribute exactly at
frequency $\Omega=2\omega_p$; no approximation is required. And
indeed, for the results discussed here, not even from the excitation
of the bound biexciton state have we found a sizable contribution
from non-Markovian (quantum memory) effects. For monochromatic cw
pump excitation and with the ansatz
$E^\ME_{0}(t)=\tilde{E}^\ME_{0}(t)\mathrm{e}^{-i\omega_pt}$,
$E^\ME_{2\mathbf{k}_p}(t)=\tilde{E}^\ME_{2\mathbf{k}_p}(t)\mathrm{e}^{-i\omega_pt}$,
$p^\XY_{0}(t)=\tilde{p}^\XY_{0}(t)\mathrm{e}^{-i\omega_pt}$,
$p^\XY_{2\mathbf{k}_p}(t)=\tilde{p}^\XY_{2\mathbf{k}_p}(t)\mathrm{e}^{-i\omega_pt}$
and
$b_{\mathbf{k}_p}(t)=\tilde{b}_{\mathbf{k}_p}(t)\mathrm{e}^{-i2\omega_pt}$,
$b_{3\mathbf{k}_p}(t)=\tilde{b}_{3\mathbf{k}_p}(t)\mathrm{e}^{-i2\omega_pt}$
the coupled probe and FWM dynamics can be written in the form
\begin{align}\label{probeFWMdyn}
\hbar\dot{\tilde{p}}(t)=M\tilde{p}(t)\,.
\end{align}
The vector
\begin{multline*}
\tilde{p}(t)=[\tilde{E}_0^\TE(t),\tilde{E}_{2\mathbf{k}_p}^{\TE\ast}(t),\tilde{E}_{0}^{\TM}(t),\tilde{E}_{2\mathbf{k}_p}^{\TM\ast}(t),
\\
\tilde{p}_{0}^{\Y}(t),\tilde{p}_{2\mathbf{k}_p}^{\Y\ast}(t),\tilde{p}_{0}^{\X}(t),\tilde{p}_{2\mathbf{k}_p}^{\X\ast}(t),\tilde{b}_{\mathbf{k}_p}(t),\tilde{b}^\ast_{3\mathbf{k}_p}(t)]^T\,,
\end{multline*}
groups field and polarization amplitude variables together. $M$ is a
time-independent matrix where all system parameters and the
steady-state pump polarization amplitude (Figs.~\ref{Fig2} and
\ref{Fig3}) and the corresponding pump field in the cavity modes
parametrically enter the analysis. For excitation with a linearly
polarized pump either exciting the TE or the TM mode, the matrix $M$
is block-diagonal for the components of probe and FWM parallel
(co-linear configuration) and perpendicular (cross-linear
configuration) to the pump's vectorial polarization state. Then for
each pump polarization state (TE or TM, respectively), the
$10\times10$ matrix $M$ can be decomposed into the $6\times6$ block
$M^\probepump_{\parallel}$ with $\sigma_s=\sigma_p$ and the
$4\times4$ block $M^\probepump_\bot$ with $\sigma_s\neq\sigma_p$
which describe the dynamics of the coupled variables in the vectors
$\tilde{p}^\probe_\parallel(t)=[\tilde{E}_0^\probe(t),\tilde{E}_{2\mathbf{k}_p}^{\probe\ast}(t),\tilde{p}_{0}^{\probe}(t),\tilde{p}_{2\mathbf{k}_p}^{\probe\ast}(t),\tilde{b}_{\mathbf{k}_p}(t),\tilde{b}^\ast_{3\mathbf{k}_p}(t)]^T$
with $\sigma_s=\sigma_p$ for the co-linear configurations and
$\tilde{p}^\probe_\bot(t)=[\tilde{E}_0^\probe(t),\tilde{E}_{2\mathbf{k}_p}^{\probe\ast}(t),\tilde{p}_{0}^{\probe}(t),\tilde{p}_{2\mathbf{k}_p}^{\probe\ast}(t)]^T$
with $\sigma_s\neq\sigma_p$ for the cross-linear configurations,
respectively. The indices $\probe$ and $\pump$ relate to the cavity
modes $\TE$ and $\TM$, or to the corresponding excitonic
polarization amplitude components X and Y that are excited by the
fields in these modes.  Note, that for the above-described
excitation situation (pump pulse propagating along the $x$ axis and
linearly polarized excitation in TE or TM mode) the $x$ component of
the polarization amplitude is exclusively excited by fields in the
TM mode and the $y$ component by fields in the TE mode. The matrices
$M^\probepump_{\parallel}$ and $M^\probepump_{\bot}$ can be derived
from Eqs.~(\ref{fieldmotion}) and (\ref{pumpmotion}) and take the
following form:
\begin{multline}\label{Mperp}
M^\probepump_{\bot}= \left(
\begin{array}{cccc}
 h_0^{\probe} & 0 & iV_0^{\probe} & 0   \\
 0 & h^{\probe\ast}_{2{k}_p} & 0 & \frac{1}{i}V_{2{k}_p}^{\probe\ast}   \\
 V_{0,\text{eff}}^{\probepump} & 0 & a_{0,\bot}^{\pump} & b^{\pump}_\bot   \\
 0 & V_{2{k}_p,\text{eff}}^{\probepump\ast} & b^{\pump\ast}_\bot &
 a_{2{k}_p,\bot}^{\pump\ast}
\end{array}
\right)\,,
\end{multline}
\begin{widetext}
\begin{align}\label{Mpar}
M^\probepump_{\parallel}=\left(
\begin{array}{cccccc}
 h^\probe_0 & 0 & iV_0^\probe & 0 & 0 & 0  \\
 0 & h^{\probe\ast}_{2{k}_p} & 0 & \frac{1}{i}V_{2{k}_p}^{\probe\ast} & 0 & 0  \\
 V_{0,\text{eff}}^{\probepump} & 0 & a_{0,\parallel}^{\pump} & b_\parallel^{\pump} & C^{\pump} & 0  \\
 0 & V_{2{k}_p,\text{eff}}^{\probepump\ast} & b_\parallel^{\pump\ast} & a_{2{k}_p,\parallel}^{\pump\ast} & 0 & C^{\pump\ast}  \\
 0 & 0 & -\frac{1}{2}C^{\pump\ast} & 0 & B_{0,{k}_p} & 0  \\
 0 & 0 & 0 & -\frac{1}{2}C^{\pump} & 0 & B^\ast_{{k}_p,2{k}_p}
\end{array}
\right)\,.
\end{align}
\end{widetext}
The time-independent coefficients are defined as:
\begin{align*}
h_{{k}}^\probe=\frac{1}{i}(\hbar\omega_k^\probe-\hbar\omega_p-i\gamma_c)\,,
\end{align*}
\begin{align*}
V_{k,\text{eff}}^\probepump=iV_k^\probe\Big(1-\tilde{A}|\tilde{p}_{\mathbf{k}_p}^\pump|^2\Big)\,,
\end{align*}
\begin{align*}
a_{k,i}^\pump=&\frac{1}{i}\Big[-\Delta\varepsilon_k-i\gamma_x+\tilde{A}V_{k_p}^\pump\tilde{p}^{\pump\ast}_{\mathbf{k}_p}\tilde{E}^{\pump}_{\mathbf{k}_p}
\\ &+\big(\tilde{\mathcal{T}}^{++}(2\omega_p)+\delta_{i,\parallel}\tilde{\mathcal{T}}^{+-}_{\text{cont}}(2\omega_p)\big)|\tilde{p}^\pump_{\mathbf{k}_p}|^2\Big]\,,
\end{align*}
\begin{align*}
b^\pump_{i}=&\frac{1}{i}\Big[(-1)^{\delta_{i,\bot}}\tilde{A}V_{k_p}^\pump\tilde{p}^{\pump}_{\mathbf{k}_p}\tilde{E}^\pump_{\mathbf{k}_p}
\\ &+\frac{1}{2}\big((-1)^{\delta_{i,\bot}}\tilde{\mathcal{T}}^{++}(2\omega_p)+\tilde{\mathcal{T}}^{+-}(2\omega_p)\big)\tilde{p}_{\mathbf{k}_p}^{\pump^2}\Big]\,,
\end{align*}
\begin{align*}
B_{k_1,k_2}=\frac{1}{i}(-\Delta\varepsilon_{k_1}-\Delta\varepsilon_{k_2}-2i\gamma_x-E^{xx}_b)\,,
\end{align*}
\begin{align*}
C^\pump=\frac{1}{i}(C_{xx}\tilde{p}_{\mathbf{k}_p}^{\pump\ast})\,,
\end{align*}
\begin{align*}
C_{xx}=\Big(\Big[\sum_{\mathbf{q}}W_{xx}^{-\dag}(\mathbf{q},0)\zeta(\mathbf{q})\Big]\Big[\sum_{\mathbf{q}'}\zeta^\dag(\mathbf{q}')W_{xx}^{-}(\mathbf{q}',0)\Big]\Big)^{\frac{1}{2}}\,.
\end{align*}
$\tilde{\mathcal{T}}^{++}(\Omega)$ and
$\tilde{\mathcal{T}}^{+-}(\Omega)$ are the Fourier transformed
correlation kernels as shown in Fig.~\ref{FigTmatrices}(b) and
defined by Eqs.~(27) and (32) of Ref.~\onlinecite{Takayama2002}. The
coupling strength $C_{xx}\approx0.54\,E_b^xa_0^x$ of the excitonic
polarization amplitudes to the bound biexciton amplitude is given by
the biexciton ground state wave function $\zeta(\mathbf{q})$ in the
electron spin singlet configuration and the corresponding
two-exciton Coulomb interaction matrix element
$W_{xx}^{-}(\mathbf{q},0)$, both as defined in Eqs.~(14) and (24) of
Ref.~\onlinecite{Takayama2002}.

For steady-state monochromatic pump excitation, the linear stability
analysis formulated in this section gives us comprehensive
information about growth (real-part of the eigenvalues of $M$) and
frequency (imaginary-part of the eigenvalues of $M$) of the
polariton modes in the probe and FWM directions, after the initial
external driving pulse (seed) in the probe direction is gone.
Information how the different eigenmodes contribute to the
polarization amplitudes and fields with different in-plane momenta
($0$ or $2\mathbf{k}_p$) can be obtained from the eigenvectors of
the matrices $M$. If at least one of the eigenvalues $\lambda_{i}$
of the matrices $M$ fulfills $\mathrm{Re}\{\lambda_{i}\}>0$, the
system is unstable. An arbitrarily small seed of $p^\XY_{0}$ or
$p^\XY_{2\mathbf{k}_p}$ ($\X$ or $\Y$, depending on which subspace
shows an unstable dynamics) or in the corresponding cavity modes
would grow exponentially until the matrix $M$ ceases to describe the
system correctly.

In addition to the strict steady-state analysis (regarding the pump
excitation), the general information obtained from the stability
analysis can -- to a large extent -- be carried over to the
discussion of pump-probe experiments with finite pulse lengths. From
the linear stability analysis reasonable predictions for growth
rates of probe and FWM signals and thus for polarization rotations
can be made as long as no external probe pulse significantly drives
the probe polariton dynamics during the period of amplification.
Furthermore, for interpretation of pulsed experiments based on the
linear stability analysis, the pump pulse must be spectrally
sufficiently narrow and pump and probe must have significant
temporal overlap. After a sufficiently long period of time that
particular eigenmode of $M$ corresponding to the eigenvalue with the
largest real part will dominate the overall outgoing signal in probe
and FWM directions. In a pump-probe experiment an incoming probe
pulse in this particular mode will be most efficiently amplified, or
for steady-state pump excitation without an incoming probe,
fluctuations in this mode (serving as a seed) will grow most
efficiently over time and dominate the signal in probe and FWM
direction after a sufficiently long growth period.

\section{Results and Discussion}

\begin{figure}[t]
\begin{center}
\includegraphics[scale=0.98]{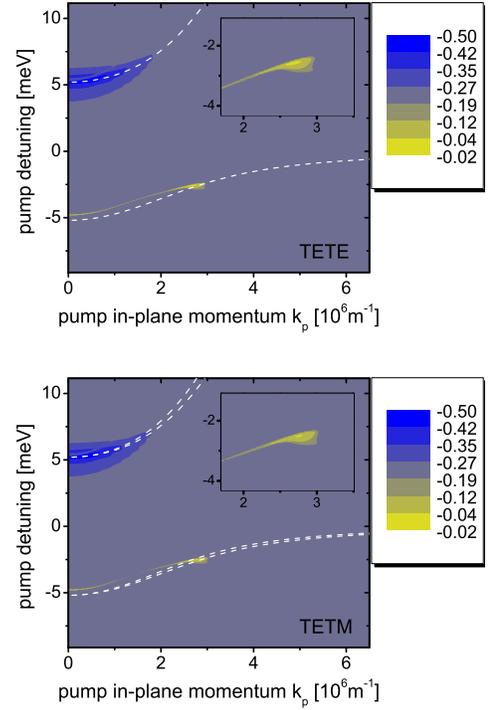}
\end{center}
\caption{\label{Fig4} (color online) Same as the two lower panels in
Fig.~\ref{Fig2} but without exciton-exciton scattering in the $+-$
channel ($\tilde{\mathcal{T}}^{+-}\equiv0$).}
\end{figure}

Without excitonic correlations (and neglecting the TE-TM-splitting
of the cavity modes) the equations of motion for the two circular
polarization state channels, $+$ and $-$, are decoupled. In this
case, for excitation with a linearly polarized pump, where equal
densities of polaritons are excited in these two different
polarization state channels, the incoming (seed in our linear
stability analysis) and outgoing probe are always in the same
vectorial polarization state, in the stable (all
$\mathrm{Re}\{\lambda\}<0$) as well as in the unstable (at least one
$\mathrm{Re}\{\lambda\}>0$) regime. Neglecting the TE-TM cavity-mode
splitting, for the pump in a linear polarization state, only the
spin-dependent excitonic correlations can give rise to a rotation of
the outgoing probe signal's vectorial polarization state relative to
the incoming one's. The actual fraction of polaritons that is
scattered into the probe and FWM directions in a vectorial
polarization state perpendicular or parallel to the pump's vectorial
polarization state, respectively, strongly depends on the excitonic
correlations in the $+-$ polarization state channel (included in
$\mathcal{T}^{+-}$).\cite{Keeling2007} As discussed in the previous
section, for a linearly polarized (either TE or TM) pump, all the
eigenmodes [eigenvectors of the matrix $M$ in
Eq.~(\ref{probeFWMdyn})] for the probe and FWM dynamics are either
polarized parallel (co-linear configuration) or perpendicular
(cross-linear configuration) to the pump, even when both excitonic
correlations and cavity-mode splitting are included. However, either
excitonic correlations alone or the cavity-mode splitting alone
(when the polariton scattering is mediated by Hartree-Fock Coulomb
interaction) is sufficient to give different probe and FWM dynamics
in the two different polarization state configurations.

Since the above-listed effects lead to a difference in the growth
rates of the modes polarized parallel or perpendicular to the pump,
for an arbitrarily polarized probe the two different vectorial
polarization state components (parallel or perpendicular) will grow
differently over time. The stronger the amplification of the probe
and the longer the amplification duration, the more the fractions of
the probe in those modes that exhibit the fastest exponential
growth, will dominate the outgoing probe and FWM signals. Thus, for
strong amplification, the growth rates of the fastest growing modes
in the two polarization states (parallel and perpendicular)
ultimately determine the rotation in the probe's vectorial
polarization state. Studying these growth rates also answers the
question about the preferred mode for the growth of probe
fluctuations over time, when no incoming probe is present.
Figures~\ref{Fig2} and \ref{Fig3} show these growth rates -- the
real part of the eigenvalue of $M$ with the largest real part in
each case -- for a fixed pump intensity for the different
polarization state configurations (co-linear or cross-linear) for
pump excitation of TE (Fig.~\ref{Fig2}) and TM (Fig.~\ref{Fig3})
mode, respectively. The results show that pumping either TE or TM
mode does not significantly influence the overall result regarding
the maximum growth rates in the co-linear or cross-linear
configuration; merely a small change in the optimum pump momentum
and frequency is observed. However, a significant difference between
co-linear and cross-linear configuration is found for pumping close
to the inflection point of the LPB (the so-called ``magic angle'')
where phase-matching is best fulfilled so that triply-resonant
(resonant for the pump excitation and at an angle so that the
dispersions in probe and FWM directions allow for phase-matched
scattering of pairs of pump-excited polaritons into these two
directions) amplification of the polaritons can occur. For the
chosen intensity, close to the instability threshold, and for pump
excitation under the ``magic angle'' we are in a regime where
instability ($\text{Re}\{\lambda\}>0$) and corresponding exponential
signal growth is only found in the cross-linear configurations
(TE-TM and TM-TE) while all the modes in the co-linear
configurations (TE-TE and TM-TM) are exponentially decaying. In this
regime, for an arbitrarily polarized probe, only that component
polarized perpendicular to the pump is exponentially growing over
time and thus only this component experiences significant
amplification. To isolate the mechanism that leads to this striking
difference in the polarization state configurations, Fig.~\ref{Fig4}
shows results for both configurations in Fig.~\ref{Fig2} but without
taking into account the correlations in the $+-$ channel
($\tilde{\mathcal{T}}^{+-}\equiv0$). Without these correlations the
results for the two configurations almost look alike; we find only a
small difference in the growth rates caused by the TE-TM cavity-mode
splitting. Note that without the correlations in the $+-$ channel in
both configurations the instability threshold is not reached for the
same pump intensity as used in Figs.~\ref{Fig2} and \ref{Fig3}.

In the co-circular ($++$) excitation configuration it was found
earlier\cite{Savasta2003} that excitonic correlations in the $++$
channel may considerably reduce the maximum growth rates in the
polariton amplification. In the $++$ channel the driving mechanism
for the instabilities, the phase-conjugate feedback, is weakened by
the two-exciton correlations. Additionally, but for large negative
detuning less important, correlations in the $++$ channel give rise
to pump-induced EID, also reducing the exponential growth rate over
time. Figures~\ref{Fig2} and \ref{Fig3} show that the correlations
in the $+-$ channel enhance the growth rate in the cross-linear
configuration and diminish it in the co-linear configuration,
compared to the results shown in Fig.~\ref{Fig4} where these
correlations are absent.

In the following we will interpret these results in terms of the
exciton-exciton scattering matrices shown in
Fig.~\ref{FigTmatrices}(b) for the different polarization state
channels of polariton-polariton scattering. For this discussion we
ignore the small PSF nonlinearities that contribute to the probe and
FWM dynamics, and concentrate on the nonlinearities in the excitonic
polarization amplitudes in Eq.~(\ref{pumpmotion}) that contribute to
the probe and FWM dynamics and therefore determine the amplification
process. Being sufficient for a qualitative understanding we discuss
all contributions in Markov approximation. Three different terms
have to be analyzed that enter the matrices $M$ in
Eqs.~(\ref{Mperp}) and (\ref{Mpar}):

(i) Excitation-induced dephasing for the excitonic component of the
polaritons (entering $M$ via $a_{k,i}^\pump$):
\begin{align}\label{EqEID}
\mathrm{Im}\Big\{\tilde{\mathcal{T}}^{++}(2\omega_p)+\delta_{i,\parallel}\tilde{\mathcal{T}}^{+-}(2\omega_p)\Big\}|\tilde{p}^\pump_\mathbf{k}|^2\,.
\end{align}
(ii) Nonlinear shifts to the effective exciton resonances in probe
and FWM direction (entering $M$ via $a_{k,i}^\pump$):
\begin{align}\label{Eqshifts}
\mathrm{Re}\Big\{\tilde{\mathcal{T}}^{++}(2\omega_p)+\delta_{i,\parallel}\tilde{\mathcal{T}}^{+-}(2\omega_p)\Big\}|\tilde{p}^\pump_\mathbf{k}|^2\,.
\end{align}
(iii) The phase-conjugate oscillation feedback for the excitonic
constituents of the polaritons that drives the instability (entering
$M$ via $b_{i}^\pump$):
\begin{align}\label{EqPCO}
\frac{1}{2}\Big((-1)^{\delta_{i,\bot}}\tilde{\mathcal{T}}^{++}(2\omega_p)+\tilde{\mathcal{T}}^{+-}(2\omega_p)\Big){\tilde{p}_{\mathbf{k}}^{{\pump^2}}}\,.
\end{align}
The index $i\in\{\bot,\parallel\}$ labels the co- and cross-linear
polarization state configurations, respectively. The maximum growth
rates in Figs.~\ref{Fig2} to \ref{Fig4} are obtained when the pump
is tuned about $3\,\mathrm{meV}$ below the exciton resonance and
close to the ``magic angle''. For this pump detuning and in Markov
approximation the two-exciton scattering matrices shown in
Fig.~\ref{FigTmatrices}(b) contribute at
$\hbar\Omega-2\varepsilon_x\approx-6\,\mathrm{meV}$. For this
detuning the $\text{Im}\{\tilde{\mathcal{T}}^{++}\}$ is much smaller
than $\text{Im}\{\tilde{\mathcal{T}}^{+-}\}$ which according to
Eq.~(\ref{EqEID}) leads to a much larger EID in the co-linear
($\parallel$) configuration. According to Eq.~(\ref{Eqshifts}) a
partial cancelation of the nonlinear energy shifts from
contributions in the $++$ and $+-$ channels is found in the
co-linear ($\parallel$) configuration only. This, however, only
slightly modifies the effective resonance frequencies (polariton
dispersions) and thus slightly changes the optimum pump momentum and
frequency. Most important for the stimulated amplification process
is the difference in the polarization state configurations that can
be seen in Eq.~(\ref{EqPCO}). Whereas for the co-linear
($\parallel$) configuration the sum of the scattering matrices in
the $++$ and $+-$ polarization state channels determines the
strength of the phase-conjugate feedback driving the instability,
for the cross-linear ($\bot$) configuration the difference of the
two is relevant. Compared to the results without correlations in the
$+-$ channel, the difference in sign in the real parts of these two
contributions [cf. Fig.~\ref{FigTmatrices}(b)] leads to strong
cancelation in Eq.~(\ref{EqPCO}) for the co-linear configuration and
to strong enhancement in the cross-linear configuration. As
previously pointed out in Ref.~\onlinecite{Renucci2005} this leads
to an imbalance in the pairwise scattering of polaritons into the
probe and FWM directions with polarization parallel or perpendicular
to the pump. From our results we conclude that this does not
necessarily lead to an overall rotation of the vectorial
polarization state by $90^\circ$ in contrast to the conclusions in
Ref.~\onlinecite{Krizhanovskii2006}. Based on our explicit treatment
of the bound biexciton, we also find that in the studied system the
scattering matrix element $\mathcal{T}^{+-}$ that drives the
amplification process by scattering of polaritons with opposite
spins is not small compared to $\mathcal{T}^{++}$. We note that this
is in contrast to the case reported in
Ref.~\onlinecite{Kavokin2005}. Our results indicate that in the cw
regime for a linearly polarized pump and close to threshold,
spontaneous fluctuations preferably grow in the cross-linear
polarization configuration. This is in agreement with recent
observations for a slightly different system and excitation
geometry.\cite{Romanelli2007}

The above discussion has a very general character. The actual
cancelation of the different contributions in
Eqs.~(\ref{EqEID})-(\ref{EqPCO}) depends quantitatively on
parameters such as the coupling strength of the cavity-modes to the
excitonic polarization amplitudes. The general trend follows from
the frequency dependence of the two-exciton scattering matrices as
shown in Fig.~\ref{FigTmatrices}(b): The stronger the cavity-mode
exciton coupling (shifting the ``magic angle'' to larger negative
detuning) the less pronounced the role of $\mathcal{T}^{+-}$ will
be. However, especially close to threshold even a small difference
in growth rates can be crucial, and in this regime even a small
$\mathcal{T}^{+-}$ contribution can play a major role for the
analysis of vectorial polarization state rotations. Although for the
system studied here it was found to be almost insignificant, the
overall role of the TE-TM cavity-mode splitting can depend on
parameters and excitation conditions, too. Furthermore the
importance of a TE-TM cavity-mode splitting can be different in
other systems such as the quasi-one-dimensional microcavity system
studied in Ref.~\onlinecite{Dasbach2005}.

Although not relevant for the amplification of polaritons on the
LPB, we finally note that Figs.~\ref{Fig2} to \ref{Fig4} show strong
pump-induced EID for excitation on the UPB. Although the two-exciton
scattering matrices evaluated in the 1s approximation are not
quantitatively accurate in this energy region, even in a theory
including only coherent excitations this additional EID would likely
inhibit the observation of any instability in the UPB (previously
discussed, e.g., in Ref.~\onlinecite{Ciuti2004}) in analogy to the
situation for positive pump detuning in a single
QW\cite{Schumacher2006}.

\section{Conclusions}

Based on a microscopic many-particle theory we have investigated the
influence of excitonic correlations on the vectorial polarization
state characteristics of the parametric amplification of polaritons
in semiconductor microcavities. By means of a linear stability
analysis it has been analyzed how a linearly polarized pump can
induce a polarization state anisotropy in an otherwise perfectly
isotropic microcavity system. For the discussion of this effect we
take advantage of our theoretical approach which -- in contrast to
previous
models\cite{Eastham2003,Kavokin2003,Kavokin2005,Renucci2005} -- is
based on microscopically calculated exciton-exciton scattering
matrix elements. Accounting for all coherent correlations between
two excitons, these matrix elements determine the nonlinear
cavity-polariton dynamics in the probe and FWM directions in the
amplification regime.

A previous study\cite{Savasta2003} found that excitonic correlations
weaken the polariton amplification for co-circular pump-probe
excitation. We confirm these findings and additionally investigate
the effects of correlations on the polariton amplification in
linearly polarized pump-probe configurations. We find that
scattering contributions of the excitonic components of polaritons
with opposite spins can strongly diminish the driving force for the
amplification, the phase-conjugate coupling, in the co-linear (TE-TE
or TM-TM) pump-probe polarization configuration and strongly enhance
it in the cross-linear (TE-TM or TM-TE) configuration. In the
spectral region where instability occurs the scattering of
polaritons with opposite spins is dominated by the virtual formation
of bound biexcitons.

In or close to the unstable regime this polarization state
anisotropy has the potential to alter the polarization state of an
amplified probe pulse compared to the incident probe. If the
incoming probe has both components polarized parallel and
perpendicular to the pump's polarization state, in general the
maximum growth rate for these two components over time is not the
same. Then after a certain growth period the amplification of these
two polarization state components will be different, and hence the
vectorial polarization state of the outgoing probe is rotated
compared to the incoming one's. Since the probe component polarized
perpendicular to the pump's polarization vector grows faster over
time than the parallel component, the probe is always rotated toward
the ``preferred'' cross-linear configuration. However, the overall
rotation in the vectorial polarization state depends on both the
duration of amplification and the difference in the growth rates in
the two polarization state channels.

Finally, we note that in a situation where steady-state pump
excitation brings only the cross-linear configuration above the
unstable amplification threshold, without an incoming probe, only
spontaneous fluctuations in the cross-linear polarization state
channel will be amplified and thus observed as a finite signal in
probe and FWM direction.

\begin{acknowledgments}

This work has been financially supported by ONR, DARPA, JSOP.
\mbox{S. Schumacher} gratefully acknowledges financial support by
the Deutsche Forschungsgemeinschaft (DFG, project No.
SCHU~1980/3-1).

\end{acknowledgments}

\appendix

\section{Nonlinear pump equation}\label{AppA}

The linear stability analysis of probe and FWM dynamics in
Sec.~\ref{Secstabana} is done for monochromatic cw pump excitation.
The stationary pump field inside the cavity, $E_{\mathbf{k}_p}^\ME$,
and pump-induced polarization amplitude in the QW,
$p_{\mathbf{k}_p}^\XY$, which enter the matrix $M$ in
Eq.~(\ref{probeFWMdyn}) are needed to analyze the probe and FWM
dynamics. In this work we have considered pump excitation with a
linearly polarized pump in the TM or TE cavity mode with in-plane
momentum $\mathbf{k}_p$ along a certain axis, here, without loss of
generality the $x$ axis, i.e., $\varphi_{\mathbf{k}_p}=0$. Seeking a
steady-state solution for the pump-induced polarization amplitude
$p_{\mathbf{k}_p}^\XY$ we use the ansatz
$E_{\mathbf{k}_p,\text{inc}}^{\ME}=\tilde{E}_{\mathbf{k}_p,\text{inc}}^{\ME}\mathrm{e}^{-i\omega_pt}$,
$E^\ME_{\mathbf{k}_p}=\tilde{E}^\ME_{\mathbf{k}_p}\mathrm{e}^{-i\omega_pt}$
for the incoming pump field and the field in the excited cavity
mode, respectively, and
$p^\XY_{\mathbf{k}_p}=\tilde{p}^\XY_{\mathbf{k}_p}\mathrm{e}^{-i\omega_pt}$
for the polarization amplitude, with
$\dot{\tilde{E}}_{\mathbf{k}_p,\text{inc}}^{\ME}=\dot{\tilde{E}}^\ME_{\mathbf{k}_p}=\dot{\tilde{p}}^\XY_{\mathbf{k}_p}=0$.
Then from Eq.~(\ref{fieldmotion}) it follows that the electric field
in the cavity mode for in-plane momentum $\mathbf{k}_p$ is given by:
\begin{align}\label{fieldmono}
\tilde{E}^\ME_{\mathbf{k}_p}=\frac{-V_{k_p}^\ME\tilde{p}^\XY_{\mathbf{k}_p}+i\hbar
t_c\tilde{E}_{\mathbf{k}_p,\text{inc}}^{\ME}}{\hbar\omega_p-\hbar\omega_{k_p}^\ME+i\gamma_c}\,.
\end{align}
For unidirectional light propagation, i.e., by removing all sums in
Eq.~(\ref{pumpmotion}), taking all field and polarization amplitude
variables at momentum $\mathbf{k}_p$, and replacing the pump field
by Eq.~(\ref{fieldmono}), a cubic equation of the form
\begin{align}\label{cubicpump}
0=a_0+a_1\big|\tilde{p}^\XY_{\mathbf{k}_p}\big|^2+a_2\big|\tilde{p}^\XY_{\mathbf{k}_p}\big|^4+\big|\tilde{p}^\XY_{\mathbf{k}_p}\big|^6
\end{align}
can be derived. This equation determines the monochromatic solutions
for each pump frequency $\omega_p$ and incoming pump intensity
($\sim \big|E_{\mathbf{k}_p,\text{inc}}^\ME\big|^2$). The
coefficients in Eq.~(\ref{cubicpump}) are
\begin{align*}
a_0=&-\frac{1}{|\tilde{T}|^2}\frac{\hbar^2t_c^2{V_{k_p}^\ME}^2|\tilde{E}_{\mathbf{k}_p,\text{inc}}^{\ME}|^2}{(\hbar\omega_p-\hbar\omega_{k_p}^\ME)^2+\gamma_c^2}\,,
\end{align*}
\begin{align*}
a_1=&\frac{1}{|\tilde{T}|^2}\Big(|\tilde{\varepsilon}|^2+\frac{2\hbar^2t_c^2{V_{k_p}^\ME}^2\tilde{A}|\tilde{E}_{\mathbf{k}_p,\text{inc}}^{\ME}|^2}{(\hbar\omega_p-\hbar\omega_{k_p}^\ME)^2+\gamma_c^2}\Big)\,,
\end{align*}
and
\begin{align*}
a_2=&\frac{2}{|\tilde{T}|^2}\Big(\mathrm{Re}\{\tilde{\varepsilon}\}\mathrm{Re}\{\tilde{T}\}+\mathrm{Im}\{\tilde{\varepsilon}\}\mathrm{Im}\{\tilde{T}\}
\\
&-\frac{\hbar^2t_c^2{V_{k_p}^\ME}^2\tilde{A}^2|\tilde{E}_{\mathbf{k}_p,\text{inc}}^{\ME}|^2}{(\hbar\omega_p-\hbar\omega_{k_p}^\ME)^2+\gamma_c^2}\Big)\,,
\end{align*}
with the definitions
\begin{align*}
\tilde{\varepsilon}=\hbar\omega^x_{k_p}-\hbar\omega_p-i\gamma_x+\frac{{V_{k_p}^\ME}^2}{\hbar\omega_p-\hbar\omega_{k_p}^\ME+i\gamma_c}\,,
\\ \tilde{T}=\frac{1}{2}(\tilde{\mathcal{T}}^{++}+\tilde{\mathcal{T}}^{+-})-\frac{{V_{k_p}^\ME}^2\tilde{A}}{\hbar\omega_p-\hbar\omega_{k_p}^\ME+i\gamma_c}\,.
\end{align*}
Being a cubic equation in $|\tilde{p}^\XY_{\mathbf{k}_p}|^2$,
depending on the coefficients $a_0,a_1,a_2$, Eq.~(\ref{cubicpump})
can have either one or three real-valued solutions. These solutions
are given, e.g., in Ref.~\onlinecite{Abramowitz1972}.

Equation~(\ref{cubicpump}) only determines the magnitude of the
pump-induced polarization amplitude $|\tilde{p}^\XY_{\mathbf{k}_p}|$
and contains no information about the different phases of the
polarization amplitude and the incoming field. We define a phase
$\phi$ according to
$\tilde{E}_{\mathbf{k}_p,\text{inc}}^{\ME}\cdot\tilde{p}^\XY_{\mathbf{k}_p}=|\tilde{E}_{\mathbf{k}_p,\text{inc}}^{\ME}||\tilde{p}^\XY_{\mathbf{k}_p}|\mathrm{e}^{i\phi}$.
This phase $\phi$ is required to determine the field in the cavity
mode from Eq.~(\ref{fieldmono}) that is coupled to the polarization
amplitude. The field enters the linear stability analysis via the
PSF terms in $M$. If we choose the solution of Eq.~(\ref{cubicpump})
to be real,
$\tilde{p}^\XY_{\mathbf{k}_p}=|\tilde{p}^\XY_{\mathbf{k}_p}|$, then
the incoming field inducing this polarization amplitude is
$\tilde{E}_{\mathbf{k}_p,\text{inc}}^{\ME}=|\tilde{E}_{\mathbf{k}_p,\text{inc}}^{\ME}|\mathrm{e}^{i\phi}$.
The phase of the incoming field can be obtained from
Eq.~(\ref{pumpmotion}) and is given by
\begin{align*}
\mathrm{e}^{i\phi}=\frac{\big(\tilde{\varepsilon}|\tilde{p}^\XY_{\mathbf{k}_p}|+\tilde{T}|\tilde{p}^\XY_{\mathbf{k}_p}|^3\big)(\hbar\omega_p-\hbar\omega_{k_p}^\ME+i\gamma_c)}{i\hbar
t_cV_{k_p}^\ME\big(1-\tilde{A}|\tilde{p}^\XY_{\mathbf{k}_p}|^2\big)\big|\tilde{E}_{\mathbf{k}_p,\text{inc}}^{\ME}\big|}\,,
\end{align*}
which then also determines the phase of the field in the cavity mode
given by Eq.~(\ref{fieldmono}).

\bibliography{../../../literature}

\begin{thebibliography}{44}
\bibitem[{Cav({\natexlab{a}})}]{CavityFootnote1}
\bibinfo{note}{The background refractive index, effective electron mass, and
  effective hole mass are $n_{\text{bg}}=3.61$, $m_e=0.067\,m_0$, and
  $m_h=0.1\,m_0$, respectively. The corresponding exciton binding energy is
  $E_b^x\approx13\,\mathrm{meV}$ and the bulk Bohr-radius is
  $a_0^x\approx170\,\AA$. For the evaluation of the theory also
  $\varepsilon^x_k=1.4965\,\mathrm{eV}$, $A^{\text{PSF}}=4a_0^x\sqrt{2\pi}/7$,
  $E_b^{xx}\approx1.84\,\mathrm{meV}$, and $\gamma_x=0.25\,\mathrm{meV}$ are
  required. The cavity parameters are $\gamma_c=0.25\,\mathrm{meV}$, and
  $V_0^\EM=5.2\,\mathrm{meV}$.}
\bibitem[{Cav({\natexlab{b}})}]{CavityFootnote2}
\bibinfo{note}{Alternatively to the chosen notation, the excitonic polarization
  can also be expressed in terms of longitudinal and transverse components with
  respect to $\mathbf{k}$. This way the ``artificial'' choice of a fixed
  coordinate system, like in our case the X-Y basis, can be avoided. However,
  then the appearance of angular dependencies containing the azimuthal angle
  relative to $\mathbf{k}$ are inherent in the nonlinear terms coupling
  polarizations and fields with different in-plane momenta.}
\end{thebibliography}
\bibliographystyle{apsrev}

\end{document}